# Can A Neural Network Hear the Shape of A Drum?

Y. Zhao and M.M. Fogler

*Department of Physics, University of California San Diego, 9500 Gilman Dr, La Jolla, CA 92093*

## Abstract

We have developed a deep neural network that reconstructs the shape of a polygonal domain given the first hundred of its Laplacian eigenvalues. Having an encoder-decoder structure, the network maps input spectra to a latent space and then predicts the discretized image of the domain on a square grid. We tested this network on randomly generated pentagons. The prediction accuracy is high and the predictions obey the Laplacian scaling rule. The network recovers the continuous rotational degree of freedom beyond the symmetry of the grid. The variation of the latent variables under the scaling transformation shows they are strongly correlated with Weyl's parameters (area, perimeter, and a certain function of the angles) of the test polygons.

## I. Introduction

The title of this paper is a reference to a famous question "Can One Hear the Shape of a Drum?" [1]; that is, whether any two isospectral planar domains are necessarily isometric. Strictly speaking, the answer is negative; however, the known counter-examples [2] are rare and generally presumed to be of measure zero [3,4]. In other words, the answer is affirmative for a vast majority of shapes [5–10]. The natural follow-up question is to how to reconstruct the shape of the domain from its spectrum, i.e., from a handful of its first Dirichlet eigenvalues. We call this inverse Dirichlet problem or IDP. Existing numerical algorithms solve the IDP by iterative morphing of the embedding mesh until the eigenvalues match those of the true ground [11,12]. This method is computationally intensive and its convergence often depends on a good initial guess.

In recent years, deep neural networks (DNN) have gained popularity as a new tool for solving a variety of inverse problems. In physical science, examples include optimization of photonic crystals, [13] data analysis in microscopy and spectroscopy, [14,15] detection of quantum phases, [16–18] and subatomic particles in collision experiments. [19–21] Some studies in computer vision used DNN to solve problems conceptually similar to IDP, e.g., reconstruction of 3D geometries from 2D images or 1D vector encodings [22–25] and 2D images from text descriptions [26–28]. It is therefore may be expected that the DNN is capable of solving the IDP, at least for certain classes of 2D shapes.

In this work, we present a proof-of-principle demonstration that DNN can accurately solve the IDP for randomly generated pentagons. Additionally, we attempt to examine the properties of the function learned by this DNN. We find evidence that the DNN discovered the scaling laws of the Laplace operator and the rotational symmetry of the problem. We also find a strong correspondence between the latent variables of the DNN and the three Weyl parameters (area $\mathcal{A}$, perimeter $\mathcal{L}$, and a certain function $\mathcal{K}$ of the angles, see below).

## II. Data Generation And Network Structure

We now describe the data generation and the structure of our DNN. The vertices of the pentagons were generated in polar coordinates with polar distances uniformly sampled between 0.5 and 2. To ensure the proper size of the pentagons, the corresponding polar angles were generated such that the included angles between neighboring radii have values between $\pi/10$ and $\pi$. The associated first 100 Dirichlet eigenvalues were calculated using a standard finite-element solver (Mathematica). It is important to notice that the computed eigenvalue spectrum is invariant under any translations, rotations, and reflections of the underlying shape. However, in IDP these isometric transformations would introduce ambiguity in the DNN output. To eliminate this "gauge freedom," we did the following. First, we shifted the centroids of the pentagons to the origin. Second, we rotated the pentagons such that the vertex with the smallest inner angle is located on the positive $x$ semi-axis. Finally, the remaining reflection symmetry was integrated into the loss function to be discussed later in this section. After the above procedure, the chosen representations were coarse-grained into $41 \times 41$ binary images $I_{\text{true}}$ by dividing the square $-2 \leq x, y \leq 2$ into pixels of size $0.1 \times 0.1$. For these input images we restricted the pixel values to be binary, $p_i = 0$ or 1. On the other hand, the output of the DNN are images $I_{\text{pred}}$ that we allowed to have fractional pixel values $0 \leq q_i \leq 1$.

To quantify the similarity between the predicted and the original images we used the Jaccard index (sometimes called the IoU index), which is defined as the ratio between the intersection (I) and the union (U) of two sets. In our case, the Jaccard index $J(I_{\text{pred}}, I_{\text{true}})$ takes the form

$$J(I_{\text{pred}}, I_{\text{true}}) = \frac{\sum_i p_i q_i}{\sum_i p_i^2 + q_i^2 - p_i q_i}. \tag{1}$$

Here the summation index $i$ runs over the entire 41x41 image. The Jaccard index is a number between zero and one, the latter being reached if the images are identical, $p_i = q_i$. To combat the aforementioned "gauge freedom" problem, we chose the loss function $L$ to be

$$L = \min_{\pi \in D_4} [1 - J(\pi(I_{\text{pred}}), I_{\text{true}})], \tag{2}$$

where $\pi$ is an element in the dihedral group $D_4$ which describes the symmetry of a square.

Our DNN has the encoder-decoder structure. Networks with such a structure have been shown to handle many complicated tasks including detecting abnormal quantum phases, [29] approximating quantum state distributions, [30] and parametrizing nonlinear dynamics [31–33]. Recent studies have also shown that encoder-decoder DNNs have the ability to discover key physical parameters of the problem [34,35]. As depicted in Fig. 1(a), the input to the DNN consists of the first 100 eigenvaule spacings. The encoder portion is composed of three consecutive long-short term memory (LSTM) [36] layers of size 128 each. The latent layer directly after the encoder contains 10 neurons with linear activation. It is crucial to notice that even though seven independent parameters are enough to uniquely determine a pentagon, we still expand the latent space to increase the DNN approximation capability. The decoder starts with four dense layers of sizes 50, 512, 1024, and 3200 with LeakyReLu activation (negative slope coefficient set to 0.3 and is the same for the rest of this section). [37] Next, the output vector of size 3,200 is reshaped to 128 feature maps each of size $5 \times 5$. Four 2D Transpose Convolution (TransConv2D) layers with kernel sizes $3 \times 3$, $3 \times 3$, $3 \times 3$, and $1 \times 1$ and kernel stride lengths of 2, 2, 2, and 1, then construct the image up-sampling block. Through such a process, the number of feature maps reduces to 64, 32, 16 and 1 and the initial $5 \times 5$ image expands to the final prediction of size $41 \times 41$. The first three TransConv2D layers have the ReLu activation while the last one has sigmoid activation to constraint the predicted pixel values. Only the last TransConv2D layer utilizes zero-padding. During training, the loss function defined in Equation (2) is applied and 20% of the samples are used for validation. A total of 1,100,000 training samples are fed into the network in 11 batches. The training epoch number for each batch decreases linearly from 50 to 10. We used the Adam optimization method with an initial learning rate of $10^{-3}$ and a decay rate of $10^{-5}$. After training, a test data set containing 100,000 new samples was used to measure the prediction accuracy

## III. Results

Our results are demonstrated in Fig. 2 where we present the cumulative distribution function (CDF) of the Jaccard index loss $L$ for the test dataset. For almost all testing samples, the loss function produces values below 0.5. The mean of the entire distribution occurs at 0.069 and only 33% of the total cases have losses larger than that. To illustrate the relation between the value of the Jaccard index and the visual goodness of the DNN predictions, we provide three examples marked "good", "mediocre," and "bad", which correspond to the top 15%, mid 70%, and the bottom 10% of the CDF, respectively. The "good" prediction captures almost every detail of the true polygonal shape (boxed in red, same for other examples), with sharp boundary and only a few pixel-level discrepancies present. In the "average" prediction, the perimeter become blurry. The major inconsistency is seen in the upper right vertex separated by a green line: this structure is missing in the prediction; however, a similar shaped triangle appears on an adjacent edge. We will discuss a possible reason for this misconstruction later in this section. In the example of a "bad" prediction, only one vertex appears sharply defined while the remaining ones become smeared. We think that these failures resulted from the rareness of such instances in the training set. Overall, since our DNN captures the true geometry of the boundaries for the majority of the tested samples, we conclude that it can indeed "hear" the shape of a pentagon from its overtones.

Next, we address the question whether the DNN is capable of learning basic properties of the Dirichlet problem. We attempt to answer this question by investigating the scaling behavior of the DNN output. For example, the magnitude of the eigenvalues scales inversely proportional to the square of the linear dimension of the drum. Therefore, scaling of the DNN input (level spacings) by a factor of $S$ should make the predicted area $\mathcal{A}$ expand by a factor of $1/S$. To check if the DNN adheres to this rule, we repeated the simulations applying scaling factors $0.5 \leq S \leq 2.5$ to the previous dataset. The results are demonstrated in Fig. 3. In Fig. 3(a) we show the scaling behavior of a typical "good" prediction. For this example, the predicted polygons vary in size according to the expected scaling rule. All major features of the drum boundary are maintained in both enlarged and shrunken images. Surprisingly, the same scaling rule is also exhibited by the "bad" prediction. As demonstrated in Fig. 3(b), the DNN not only correctly boosts this faulty prediction but it also preserves all its major features. To check the area scaling quantitatively, we defined the area $\mathcal{A}$ of the predicted pentagon to be the sum of all the pixel values. We observed that it has a power-law dependence on the eigenvalue scaling factor $S$ with the exponent of $-0.93$, which is very close to the expected $-1$, see Fig. 3(c).

Another surprise is that the DNN also discovers the continuous rotational degree of freedom beyond the symmetry of the underlying square grid. As exemplified in Fig. 4(a), after a counterclockwise rotation of $150°$, a seemingly fallacious prediction with $L = 0.35$ shows a much better agreement with the true ground ($L = 0.13$). Such cases make up more than one quarter of the worst 3000 predictions. Additional examples of seemingly "bad" predictions include instances where $L$ can be greatly reduced if a suitable rotation is combined with a reflection, see Fig. 4(b). It is remarkable that these predictions exist despite being penalized by the loss function. They suggest that the DNN must have captured fundamental information associated with the geometry of the drumheads.

Frequently discussed in the context of the IDP is Weyl's formula [3,38,39] for the average number of eigenvalues below a prescribed value $E$:

$$\overline{\mathcal{N}}(E) \simeq \frac{\mathcal{A}}{4\pi}E - \frac{\mathcal{L}}{4\pi}\sqrt{E} + \mathcal{K}. \tag{3}$$

Here $\mathcal{L}$ is the perimeter of the polygon, constant $\mathcal{K}$ is given by

$$\mathcal{K} = \frac{1}{24}\sum_i \left(\frac{\pi}{\alpha_i} - \frac{\alpha_i}{\pi}\right), \tag{4}$$

and $0 < \alpha_i < 2\pi$ are the inner angles. We find evidence that the DNN discovers Weyl's expansion and stores information about the three parameters $\mathcal{A}/4\pi$, $\mathcal{L}/4\pi$, and $\mathcal{K}$ in the latent neurons. To do so we utilize the universal approximation property [40] of neural networks to establish a mapping from the latent neurons to Weyl's parameters. Specifically, we detached the encoder and the latent layer from the original DNN and connected it instead to a small latent layer analyzer network whose outputs were Weyl's parameters (or other parameters of interest, see below). To control the complexity of the mapping function, we varied the number of hidden layers in the small network but fixed the size of each layer to 50 neurons, see Fig. 1(b). We kept the output layer activation-free and hence the only source of nonlinearity in the small network was the hidden layers where the LeakyReLu activation function was applied. During training, we allowed modifications of the small network parameters only, using the mean squared error (MSE) as the loss function. In a similar way, we constructed two other networks, one predicting Cartesian coordinates of the polygon vertices and another one predicting the edge lengths and inner angles of the polygons. These more traditional ways to define shapes were included for comparison.

We used 1,100,000 samples to train the mapping from the latent space to each of these three sets of boundary descriptors. The same optimizer with identical learing rate and decay policy was applied. The testing results from 100,000 samples are demonstrated in Fig. 5(a), where the mean predicted percentage errors for every parameter set are graphed as functions of the number of the hidden layers. Notably, for Weyl's parameters, the error is as low as 7% for a simple linear mapping. When nonlinearity is later introduced, the error plateaus at 1.6%. In contrast, even equipped with the highest level of nonlinearity, the network struggles to construct the mapping functions for the remaining two sets of the descriptors. The prediction accuracies hardly surpass those of random guess. We conclude that the latent representation has the closest relation to Weyl's parameters.

At this point it still remains a question whether the combined network understands the meaning of $\mathcal{A}$, $\mathcal{L}$, and $\mathcal{K}$. Since no additional physical information, e.g., measurement units, was provided during training, it is posssible that the learned mapping treated the Weyl's parameters as unitless and hence overfitted them. To check if that is the case we again investigated the scaling behavior of the outputs. Unlike the DNN predictions of the images, the output Weyl's parameters should obey different scaling laws: $\mathcal{A}$ should scale as $1/S$, $\mathcal{L}$ as $1/\sqrt{S}$, while $\mathcal{K}$ should be scale-invariant. As demonstrated in Fig. 5(b), for a network with one hidden layer, the average $\mathcal{A}$, $\mathcal{L}$, and $\mathcal{K}$ predictions indeed follow the correct scaling laws. In this Figure we included only the samples and scaling factors $S$ for which the scaled shapes remain within the training range (for the entire data set, see Fig. S1 in the Supplementary materials). From the above results, we confirm that Weyl's parameters are stored in the latent space quasi-linearly, and so the prediction of the drum shapes by the encoder-decoder network may be aided by these parameters. Yet another evidence in favor of this interpretation is obtained by visually examining the "mediocre" predictions. As demonstrated in Fig. 6, this prediction misses the upper right corner of the pentagon, as outlined by the green triangle but includes an extra triangular region depicted in blue. These two regions, the missing one and the added one, are approximately isometric, and so the predicted image preserves Weyl's parameters of the true ground. The results from the latent layer analyzer network further validate the above analysis: the predicted $\mathcal{A}$, $\mathcal{L}$, and $\mathcal{K}$ for this example are just only 0.64%, 1.2%, and 3.2% off from the ground truth. Since the latent space is preserved as a consequence, the DNN "sees" the wrong geometry as exactly the ground truth when generating the prediction.

## IV. Conclusion

In conclusion, we have developed an encoder-decoder DNN that solves the IDP for simple shapes such as pentagons with an exceptional accuracy. We presented evidence that this DNN has learned the scaling properties of the Laplacian operator and that it stores information about Weyl's expansion parameters (area $\mathcal{A}$, perimeter $\mathcal{L}$, and inner-angle characteristic $\mathcal{K}$) in the latent representation. The DNN has discovered a continuous rotational symmetry of the Dirichlet problem beyond the limitations of the square grid. Note that the latent space must also contain information about other boundary descriptors, such as the Cartesian coordinates of the vertices. However, we found extracting these more readily interpretable parameters challenging. We think that by implementing a mutually independent latent space [34] and more extensive training, one could potentially do so as well. Since our predicted domains are visualized in pixel grids, the current DNN can be easily generalized and transferred to generate more complicated non-polygonal shapes with refined resolution.

As mentioned in the introduction, there are examples of eigenvalue spectra [2] for which the solution of the IDP is not unique, i.e., there exist non-isometric domains with identical spectra. These examples are rare (measure zero) but very interesting from the theoretical point of view [3]. We envision that an improved DNN architecture could discover new families of such non-isometric isospectral domain pairs, which may further guide human intuition in solving mathematical problems [41].

Our study was originally motivated by scanning probe optical nanoscopy, where the methods developed to tackle IDP could help to achieve super-resolution of the shapes of nanoparticles using their collective mode spectra [42]. We hope that our work would stimulate further applications of DNN to solve inverse spectral problems in this and various other fields.

*Note added*. When this paper has been completed, we learned of a related work [43] where the DNN was successfully applied to solving the IDP for 3D shapes representing human facial expressions. That work was motivated by industrial applications of computer vision techniques. No attempt to analyze the physical and methematical properties of the latent variables was made.

# Acknowledgements

We thank Yi-Zhuang You for valuable suggestions on improving the network design and Mengkun Liu and Xinzhong Chen for useful comments and discussions.

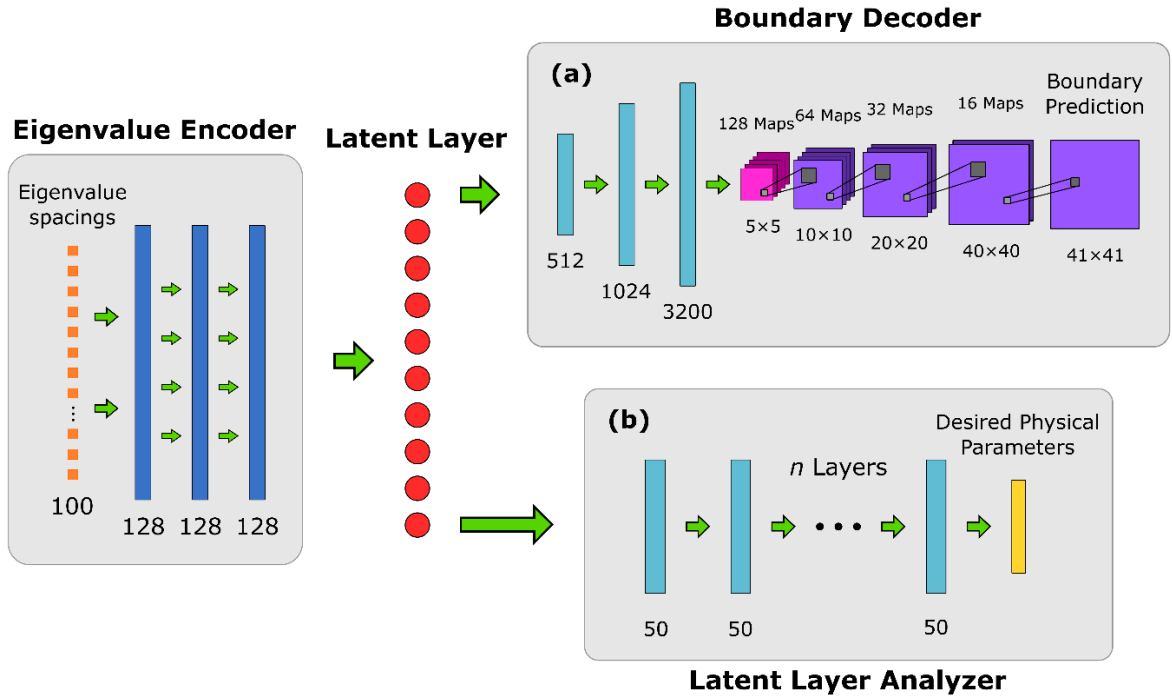

Fig 1. **Structure of the DNN developed in this study.** The full architecture contains an eigenvalue encoder, a latent layer, a boundary decoder and a latent layer analyzer. The encoder consists of three consecutive LSTM layers (dark blue) that extract information from the input eigenvalue spacings. This information is then compressed into the latent layer of 10 neurons. The decoder first expands the latent space via three dense layers (cyan), then reshapes (magenta) the result, then decodes the image using four 2D transpose convolutional layers (purple). The latent layer analyzer deciphers the latent space using $n$ consecutive hidden dense layers. **(a)** The encoder and the latent layer are connected to the decoder for boundary prediction task. Here we train all the sections. **(b)** The latent layer analyzer is connected to the encoder and latent layer for physical parameter extraction. In training, only the weights in the analyzer are adjusted.

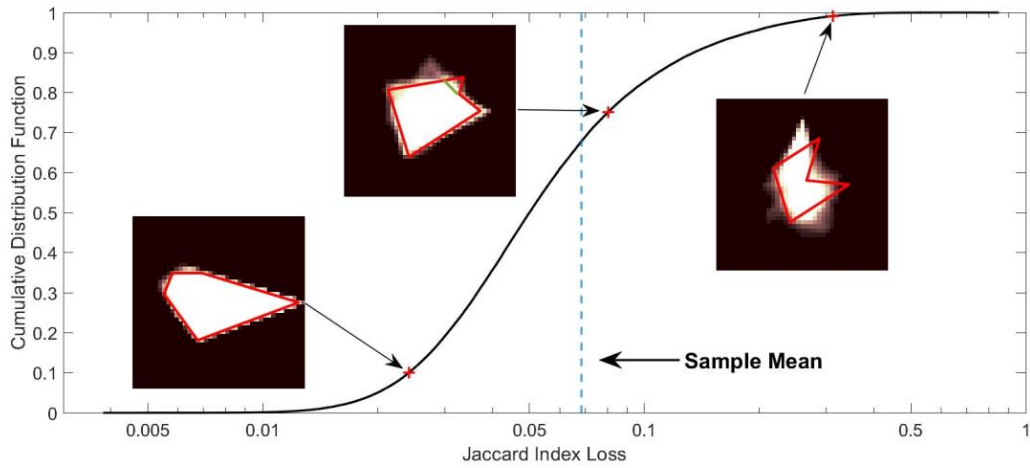

Fig 2. **Cumulative Distribution Function (CDF) of the Jaccard Index (IoU) loss function for a test dataset of 100,000 samples.** The dashed vertical line indicates the sample mean. The crosses on the curve mark typical "good", "mediocre," and **"bad"** predictions. The red polygons in the images indicate the true ground in each case.

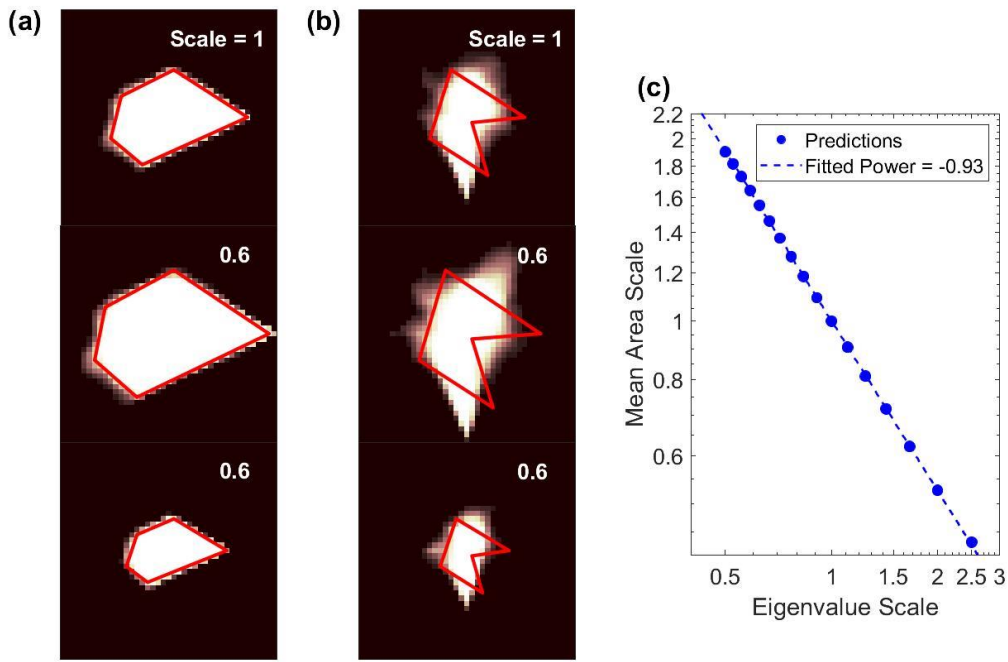

Fig 3. **The DNN discovers the scaling property of the Laplace operator.** (a) Examples of the scaling behavior of the predicted images with the corresponding eigenvalue scaling factors indicated. The ground truths are demonstrated in red for comparison. (b) The scaling law holds even for a poor prediction. (c) The predicted area as a function of the scaling factor, fitted to a power law (dashed line) with exponent $-0.93$. The statistical error is smaller than the symbol size.

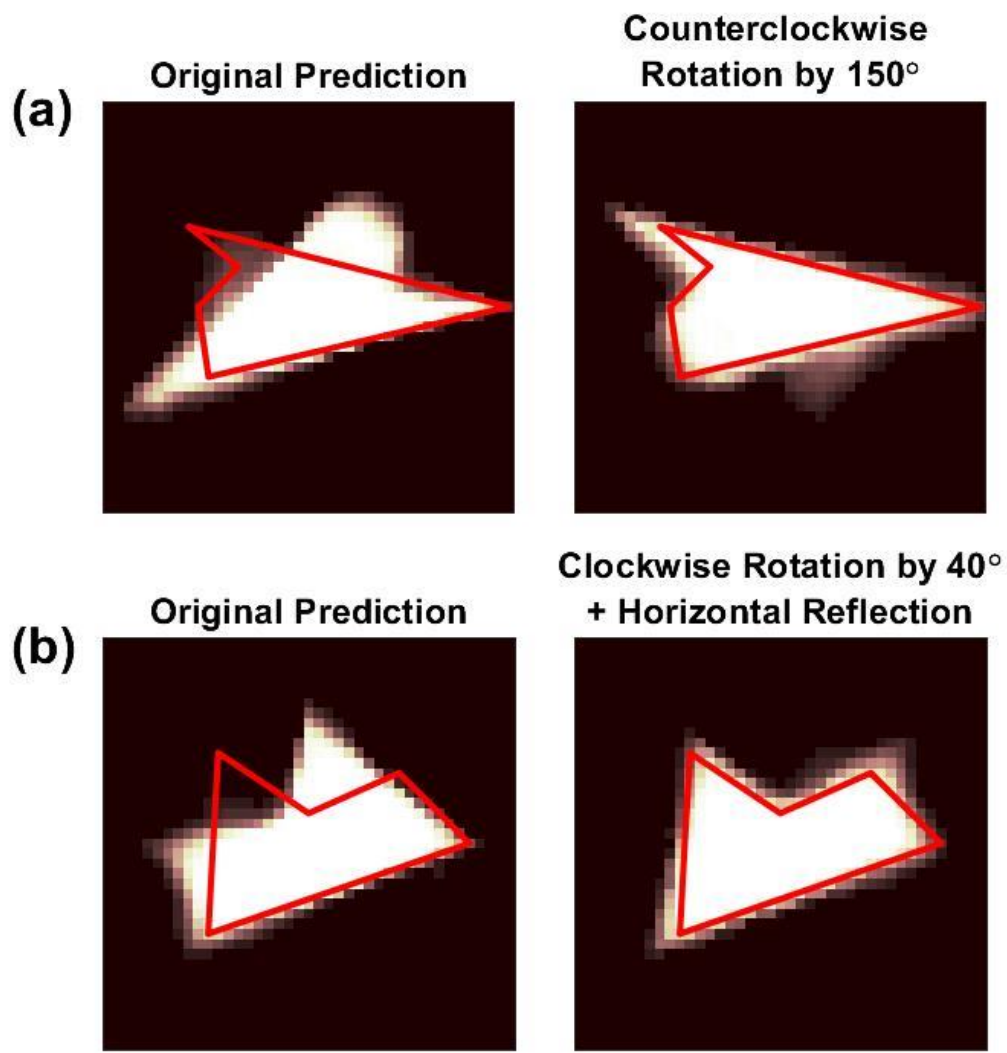

Fig 4. **The DNN discovers spatial symmetries beyond the limitations of the underlying square grid.**
(a) an example of the rotational symmetry (b) an example of a rotation combined with a reflection. The true grounds are displayed in red.

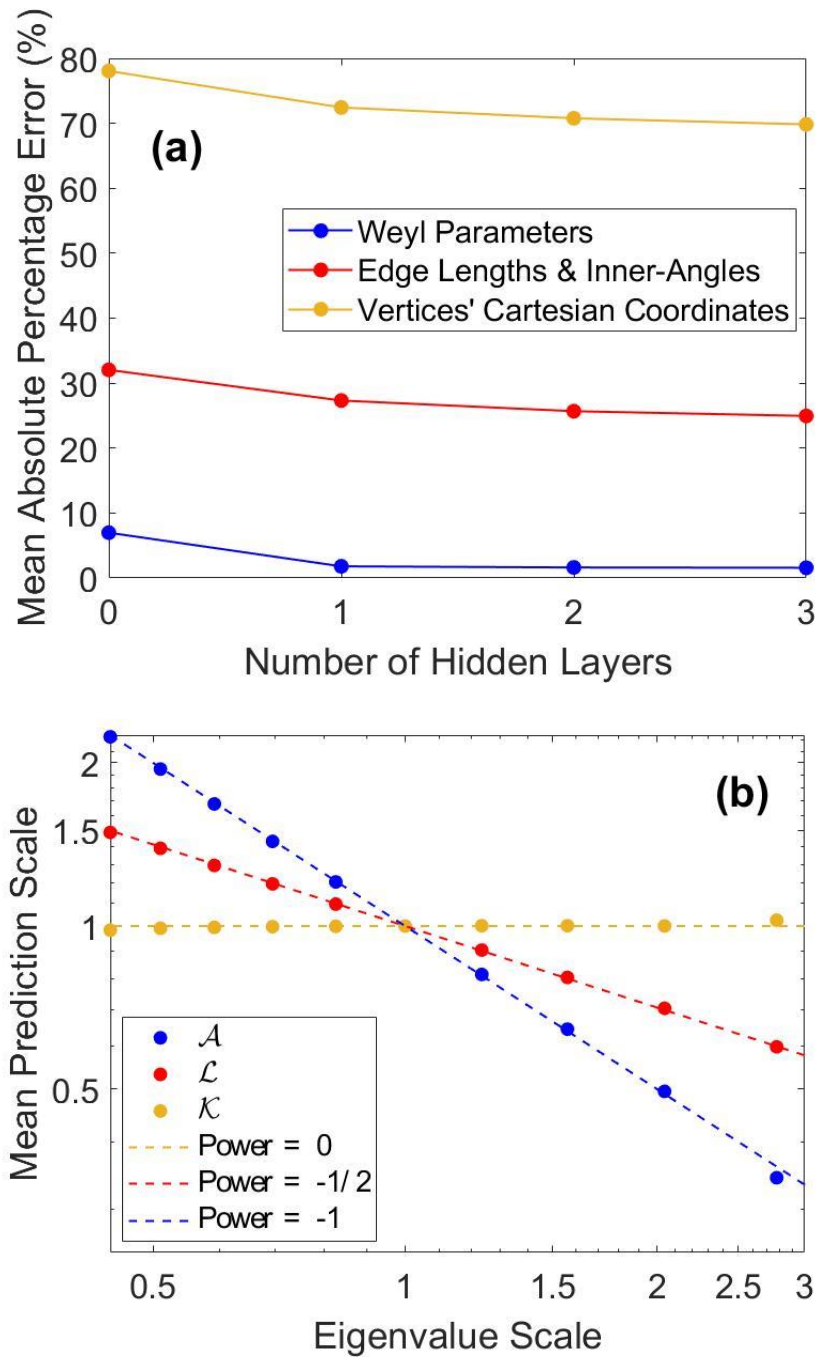

Fig 5. **Latent space analysis.** (a) The prediction accuracy for different sets of geometric parameters (Weyl's parameters, blue; the edge lengths and inner angles, red; the Cartesian coordinates of the vertices, yellow) as a function of the number of hidden layers in the analyzer. (b) Scaling behavior of the Weyl's parameters (area, red; perimeter, blue; inner angle characteristic, yellow) predicted by the analyzer with a single hidden layer. The dashed lines represent the proper scaling laws. The test set contained 10,074 samples that remained within the grid boundary in the full range of the eigenvalue scaling factors shown.

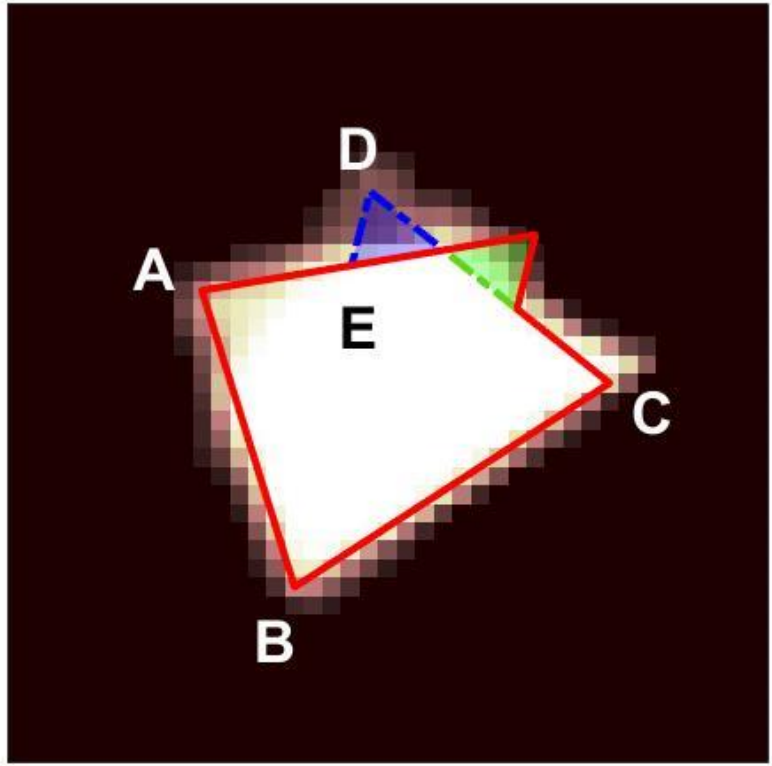

Fig 6. **Weyl's parameters preserved in a mediocre prediction.** An expanded view of the "mediocre" prediction in Fig. 2 approximated by a polygon ABCDE. The blue and green triangles are congruent. This implies that ABCDE has the same Weyl's parameters as the red polygon (the ground truth rotated by a small angle).

# Citation